\documentclass[aps,preprint,showpacs,tightenlines]{revtex4}
\usepackage{graphicx}
\usepackage{amsmath}
\usepackage{amssymb}

\begin{document}

\preprint{}
\title{Coupled Ripplon-Plasmon Modes in a Multielectron Bubble}%
\author{S.N. Klimin$^{1,*}$, V.M. Fomin$^{1,*,**}$, J. Tempere$^{1,2}$,
I.F. Silvera$^2$, and J. T. Devreese$^{1,\;\sharp}$}%
\affiliation{$^1$Theoretische Fysica van de Vaste Stoffen (TFVS),
Universiteit Antwerpen (UIA), B-2610 Antwerpen, Belgium}%
\affiliation{$^2$Lyman Laboratory of Physics, Harvard University,
Cambridge, Massachusetts 02138, USA}%

\begin{abstract}
In multielectron bubbles, the electrons form an effectively
two-dimensional layer at the inner surface of the bubble in helium. The
modes of oscillation of the bubble surface (the ripplons) are influenced
by the charge redistribution of the electrons along the surface. The
dispersion relation for these charge redistribution modes (`longitudinal
plasmons') is derived and the coupling of these modes to the ripplons is
analysed. We find that the ripplon-plasmon coupling in a multielectron
bubble differs markedly from that of electrons a flat helium surface. An
equation is presented relating the spherical harmonic components of the
charge redistribution to those of the shape deformation of the bubble.
\end{abstract}

\date{\today}
\maketitle


\section{Introduction}

Multielectron bubbles are 0.1 $\mu $m -- 100 $\mu $m sized cavities inside
liquid helium, that contain helium vapor at vapor pressure and a
nanometer-thick electron layer anchored to the surface of the bubble \cite
{VolodinJETP26}. The dependence of the radius of the bubble on the pressure
applied to the bubble \cite{TemperePRL87} allows to vary the surface density
of the spherical two-dimensional electron system in the bubble over nearly
four orders of magnitude. As such, these objects offer the prospect to study
the two-dimensional electron system on helium in regimes hitherto
inaccessible. In particular, this would allow to study the two-dimensional
electron Wigner crystal at densities not currently achievable for electrons
on a flat helium surface.

Early experimental evidence for 2D Wigner crystallization of electrons on a
liquid-He surface \cite{GrimesPRL42} relied on the detection of the coupled
plasmon-ripplon modes \cite{FisherPRL42}. In Ref. \cite{TemperePRL87}, the
ripplon modes characteristic for a bubble inside liquid helium were
calculated and investigated as a function of the pressure applied on the
bubble. In this communication we extend this investigation and take into
account the redistribution of charge along the bubble surface when the
bubble deforms. This allows us to derive the dispersion relation for the
spherical `plasmon' modes and the plasmon-ripplon coupling peculiar for the
spherical surface.

\section{Small amplitude deformations and charge redistribution}

The surface of a bubble in liquid helium can be described by a function $%
R(\Omega )$ which gives the distance of the helium surface from the center
of the bubble, in a direction given by the spherical angle $\Omega =\{\theta
,\phi \}$. In equilibrium (at zero pressure) the multielectron bubble has a
spherical surface with radius $R_{\text{b}}$. We expand the deformation away
from the spherical surface in spherical harmonics:
\begin{equation}
R(\Omega )=R_{\text{b}}+\sum_{\ell =1}^{\infty }\sum_{m=-\ell }^{\ell
}Q_{\ell m}Y_{\ell m}(\Omega ).  \label{Rexp}
\end{equation}
where $R_{\text{b}}$ is the angle-averaged radius of the bubble, $Q_{\ell m}$
is the amplitude of the spherical harmonic deformation mode indexed by $%
\{\ell ,m\}$, and $Y_{\ell m}(\Omega )$ is the corresponding spherical
harmonic. These modes of deformation are referred to as \textit{ripplons} in
analogy with the surface modes of a flat surface of liquid helium. In Ref.
\cite{TemperePRL87} the dispersion relation for the ripplons in a
multielectron bubble has been studied as a function of pressure and related
to the bubble stability. In this communication,we extend the results of \cite
{TemperePRL87} in order to take the coupling and mixing of the ripplons and
the electronic modes into account.

In a multielectron bubble, the electrons are confined to a thin (1 nm) layer
at the inner surface of the bubble. This layer is anchored to the surface of
the bubble, so that when the surface deforms, the layer conforms to the new
bubble surface. The electrons can redistribute themselves inside the
spherical layer, so that the surface density of electrons is no longer
uniform. We describe the surface density with the function $n_{S}(\Omega )$
giving the number of electrons in a solid angle $d\Omega $ around $\Omega :$%
\begin{equation}
n_{S}(\Omega )=\dfrac{N}{4\pi R_{b}^{2}}+
\sum_{\ell =1}^{\infty }\sum_{m=-\ell }^{\ell}
n_{\ell m}Y_{\ell
m}(\Omega ).  \label{nexp}
\end{equation}
This function is normalized such that $N=\int n_{S}(\Omega )R_{b}^{2}d\Omega
$ is the total number of electrons in the bubble or on the droplet. The $%
n_{\ell m}$'s represent the amplitudes of charge redistributions
corresponding to spherical harmonics $Y_{\ell m}$.

The total potential energy of the multielectron bubble or droplet can be
separated into several contributions: (i) a term from the surface tension
energy $U_{S}=\sigma S$, with $\sigma \approx 3.6\times 10^{-4}$ J/m$^{2}$
and $S$ the surface of the deformed bubble or droplet; (ii) a pressure
related term $U_{p}=pV$, with $p$ the difference in pressure outside and
inside the bubble and $V$ the volume of the bubble; (iii) a term
representing the electrostatic energy $U_{C}$ of the electron layer. In this
last term, the quantum corrections (such as the exchange contribution) to
the electrostatic energy can be neglected \cite{ShungPRB45}. The first two
terms, $U_{S}$ and $U_{p}$, have been derived in \cite{TemperePRL87}:
\begin{eqnarray}
U_{S} &=&4\pi \sigma R_{b}^{2}+\dfrac{\sigma }{2}\sum_{\ell
=1}^{\infty }\sum_{m=-\ell }^{\ell }\left( \ell ^{2}+\ell +2\right)
\left| Q_{\ell m}\right| ^{2},  \label{US} \\
U_{p} &=&\dfrac{4\pi }{3}pR_{b}^{3}+pR_{b}\sum_{\ell =1}^{\infty
}\sum_{m=-\ell }^{\ell }\left| Q_{\ell m}\right| ^{2}.  \label{Up}
\end{eqnarray}
Whereas $U_{S}$ is an exact expression, $U_{p}$ is an expansion up to second
order in $Q_{\ell m}$ and the third order term has been neglected. In what
follows, we assume small amplitude deformations and small amplitude charge
redistributions such that
\begin{eqnarray}
\sqrt{\ell (\ell +1)}\left| Q_{\ell m}\right|  &\ll &R_{b},  \label{Ransatz}
\\
\sqrt{\ell (\ell +1)}\left| n_{\ell m}\right|  &\ll &\dfrac{N}{4\pi R_{b}^{2}%
}.  \label{nansatz}
\end{eqnarray}

The electrostatic potential $V(\mathbf{r})$ of the deformed MEB with a
non-uniform surface electron density is calculated straightforwardly within
the framework of the Maxwell equations by expanding the potential inside and
outside the deformed bubble in their respective spherical decompositions and
imposing the electrostatic boundary conditions at the surface. The potential
energy associated with this electrostatic potential is given by
\begin{equation}
U_{C}=-(1/2)\int n_{S}(\Omega )V(R,\Omega )d\Omega.
\end{equation}
If we keep only terms
up to second order in both $n_{\ell m}$ and $Q_{\ell m}$, we find
\begin{eqnarray}
U_{C} &=&\frac{e^{2}N^{2}}{2\varepsilon R_{b}}+2\pi
e^{2}R_{b}^{3}\sum_{\ell =1}^{\infty }\sum_{m=-\ell }^{\ell }%
\frac{\left| n_{\ell m}\right| ^{2}}{\varepsilon _{1}\ell +\varepsilon
_{2}\left( \ell +1\right) }  \notag \\
&&-\frac{e^{2}N^{2}}{8\pi \varepsilon R_{b}^{3}}\sum_{\ell
=1}^{\infty }\sum_{m=-\ell }^{\ell }\frac{\varepsilon _{1}\ell
^{2}-\varepsilon _{2}\left( \ell +1\right) }{\varepsilon _{1}\ell
+\varepsilon _{2}\left( \ell +1\right) }\left| Q_{\ell m}\right| ^{2}  \notag
\\
&&-e^{2}N\sum_{\ell =1}^{\infty }\sum_{m=-\ell }^{\ell }%
\frac{\ell +1}{\varepsilon _{1}\ell +\varepsilon _{2}\left( \ell +1\right) }%
n_{\ell m}Q_{\ell m}^{\ast }.  \label{UC}
\end{eqnarray}
where $e$ is the electron charge, $\varepsilon _{1}$ is the dielectic
constant of the medium inside the surface, and $\varepsilon _{2}$ is the
dielectric constant of the medium outside the surface. For a multielectron
bubble, $\varepsilon _{2}=\varepsilon $ with $\varepsilon =1.0572$ the
dielectric constant of liquid helium, and $\varepsilon _{1}=1$ (if the
vapour pressure of the helium is low enough, the bubble has vacuum inside).

The ripplon contribution to the kinetic energy of the MEB is associated with
the motion of the liquid helium surface. Following a derivation of Lord
Rayleigh for oscillating liquid droplets \cite{RayleighPRS29}, we find
\begin{equation}
T_{\text{r}}=\frac{\rho R_{b}^{3}}{2}\dot{R}_{b}^{2}+\frac{\rho R_{b}^{3}}{2}%
\sum_{\ell =1}^{\infty }\sum_{m=-\ell }^{\ell }\frac{1}{\ell
+1}\left| \dot{Q}_{\ell m}\right| ^{2}.  \label{Tr}
\end{equation}
where $\rho \approx 145$ kg/m$^{3}$ is the density of liquid helium. Note
that for a bubble, $\ell +1$ appears in the denominator instead of $\ell $
for the droplet in Lord Rayleigh's treatment.

The `classical' kinetic energy of the electrons is given by
\begin{equation}
T_{\text{e}}=\dfrac{m_{\text{e}}}{2}\sum_{j=1}^{N}\mathbf{\dot{w}}^{2}(%
\mathbf{r}_{j})
\end{equation}
with $m_{\text{e}}$ the electron mass, and $\mathbf{w}(\mathbf{r}_{j})$
represents the displacement of electron $j$ out of its equilibrium position $%
\mathbf{r}_{j}$. Since we assumed (\ref{nansatz}), we can use the formula
\begin{equation}
T_{\text{e}}=\dfrac{Nm_{\text{e}}}{8\pi }\int_{\text{surface}}\mathbf{\dot{w}%
}^{2}(\mathbf{r})d^{2}\mathbf{r}
\end{equation}
and express the field of displacements $\mathbf{w}(\mathbf{r})$ as a sum of
a longitudinal field $\mathbf{w}_{\text{L}}(\mathbf{r})$ and transverse
field $\mathbf{w}_{\text{T}}(\mathbf{r}).$ We investigate the effect of
the longitudinal field, which can be written as a gradient of a scalar
potential. The (divergence-free) transverse field is not considered here
since it does not couple to the ripplons. Using
\begin{equation}
\nabla \cdot \mathbf{w_L}(\mathbf{r})=1-\dfrac{4\pi
R_{b}^{2}}{N}n_{S}(\theta,\phi)
\end{equation}
we can express the kinetic energy of the electrons as
\begin{equation}
T_{\text{e}}=\dfrac{1}{2}\sum_{\ell =1}^{\infty}
\sum_{m=-\ell }^{\ell }\dfrac{4\pi m_{\text{e}}R_{b}^{6}}{N\ell
(\ell +1)}\left| \dot{n}_{\ell m}\right| ^{2}  \label{Te}
\end{equation}
We have checked that our approach yielded, for the 3D and the 2D electron
gas, the known expressions for the plasma frequencies.

\section{Results and discussion}

The full Lagrangian of the bubble, including the ripplon modes and charge
redistribution modes, is given by substituting expressions (\ref{Tr}),(\ref
{Te}),(\ref{US}),(\ref{Up}),(\ref{UC}) in $\mathcal{L}=T_{\text{r}}+T_{\text{%
e}}-U_{S}-U_{p}-U_{C}$. The result can be brought in the following form
\begin{eqnarray}
\mathcal{L} &=&\mathcal{L}_{R}+\sum_{\ell ,m}\dfrac{M_{\ell }}{2}\left( \dot{%
Q}_{\ell m}^{2}-\omega _{r}^{2}(\ell )Q_{\ell m}^{2}\right)   \notag \\
&&+\sum_{\ell ,m}\dfrac{m_{\ell }}{2}\left( \dot{n}_{\ell m}^{2}-\omega
_{p}^{2}(\ell )n_{\ell m}^{2}\right) +\sum_{\ell ,m}\gamma _{\ell }n_{\ell
m}Q_{\ell m}.  \label{Lagrangean}
\end{eqnarray}
This is the central result of this communication. We will now proceed to
discuss the result term by term. The first term in expression (\ref
{Lagrangean}) contains the Lagrangian describing the radial motion (the
anharmonic breathing mode):
\begin{equation}
\mathcal{L}_{R}=\dfrac{\rho R_{b}^{3}}{2}\dot{R}_{b}^{2}-\dfrac{4\pi }{3}%
pR_{b}^{3}-4\pi \sigma R_{b}^{2}-\dfrac{N^{2}e^{2}}{2\varepsilon R_{b}}.
\label{LR}
\end{equation}
In what follows we assume that $\dot{R}_{b}=0$ and the bubble
radius is given by its equilibrium value.

\subsection{Ripplon modes}

The Lagrangian $\mathcal{L}$ also contains a part representing the
harmonic oscillation of the ripplon
modes, with oscillator mass term
\begin{equation}
M_{\ell }=\dfrac{\rho R_{b}^{3}}{\ell +1},  \label{mr}
\end{equation}
and bare ripplon frequency:
\begin{equation}
\omega _{r}(\ell )=\sqrt{\dfrac{1}{M_{\ell }}\left[ \sigma (\ell ^{2}+\ell
+1)+pR_{b}-\dfrac{N^{2}e^{2}}{4\pi \varepsilon R_{b}^{3}}\dfrac{\ell
^{2}-\varepsilon (\ell +1)}{\ell +\varepsilon (\ell +1)}\right] }  \label{wr}
\end{equation}
The bare ripplon frequencies and their dependence on the pressure were the
subject of \cite{TemperePRL87}. Note that, in the limit of very large
bubbles, $R_{b}\rightarrow \infty $, the dispersion relation for ripplons on
a flat helium surface \cite{Atkins} is recovered: if the momentum $k$ is
identified with $\ell /R_{b}$, then the typical $k^{3/2}$ is retrieved. Note
furthermore that, if $R_{b}$ is set equal to the equilibrium coulomb radius,
the modes $\ell =1,2$ obtain a zero frequency in agreement with the result
of Salomaa and Williams \cite{SalomaaPRL47}. For typical multielectron
bubbles with radii in the range of 1-100 $\mu $m (and numbers of electrons
of the order $10^{4}-10^{7}$) the ripplon frequencies for $\ell <1000$ lie
typically in the MHz-GHz range.

\subsection{Longitudinal plasmon modes}

Next, the Lagrangean (\ref{Lagrangean}) contains a part representing the
(small amplitude) harmonic oscillation of the classical charge
redistribution around its equilibrium. The oscillator mass term for these
oscillations is
\begin{equation}
m_{\ell }=\dfrac{4\pi m_{\text{e}}R_{b}^{6}}{N\ell (\ell +1)},  \label{mp}
\end{equation}
and the corresponding frequencies are given by
\begin{equation}
\omega _{p}(\ell )=\sqrt{\dfrac{Ne^{2}}{m_{\text{e}}R_{b}^{3}}\dfrac{\ell
(\ell +1)}{\ell +\varepsilon (\ell +1)}}.  \label{wp}
\end{equation}
In the limit of large bubbles, this frequency corresponds to that of the
longitudinal plasmon frequency of a 2D electron system on the helium
surface \cite{PlatzmanPRB10}. If the electrons form a Wigner lattice, this
is the longitudinal phonon frequency. If the momentum $k$ is identified
with $\ell /R_{b}$, then the typical $k^{1/2}$ disperson for this
longitudinal plasmon is retrieved. Based on this correspondence, we will
refer to these modes in a multielectron bubble as the `(longitudinal)
plasmon modes of the MEB' (or `longitudinal phonon modes of the MEB' if
the electrons form a Wigner lattice).

For typical multielectron bubbles with radii in the range of 1-100 $\mu $m
(and numbers of electrons of the order $10^{4}-10^{7}$) the longitudinal
plasmon frequencies for $\ell <1000$ lie typically in the GHz-THz range.
For bubbles with $N>10^{5}$, the longitudinal plasmon frequencies are
larger than the ripplon frequencies, $\omega _{p}(\ell )\gg \omega
_{r}(\ell )$.

The equations of motion for the electron charge redistribution on a
deformed bubble are given by:
\begin{eqnarray}
\dfrac{d}{dt}\dfrac{\partial \mathcal{L}}{\partial \dot{n}_{\ell m}} &=&%
\dfrac{\partial \mathcal{L}}{\partial n_{\ell m}}  \notag \\
&\Leftrightarrow &m_{\ell }\ddot{n}_{\ell m}=-m_{\ell }\omega _{p}^{2}(\ell
)n_{\ell m}+\gamma _{\ell }Q_{\ell m}.  \label{neq1}
\end{eqnarray}
When $\omega _{p}(\ell )\gg \omega _{r}(\ell )$ the electrons can
redistribute much faster on the surface of the bubble than the bubble
surface can deform. Thus, we can make an adiabatic approximation to find the
charge redistribution of the electrons on a bubble with a given deformation.
We find from (\ref{neq1}) that if the bubble surface deformation is
described by a given set of spherical deformation amplitudes $Q_{\ell m}$,
the equilibrium charge distribution on the deformed bubble must satisfy the
relation
\begin{equation}
\dfrac{n_{\ell m}}{n_{0}}=\dfrac{(\ell +1)}{2}\dfrac{Q_{\ell m}}{R_{b}},
\label{chargerule}
\end{equation}
with $n_{0}=N/(4\pi R_{b}^{2})$.

\subsection{Ripplon - longitudinal plasmon coupling}

\begin{figure}[b]
\begin{center}
\includegraphics[width=13cm]{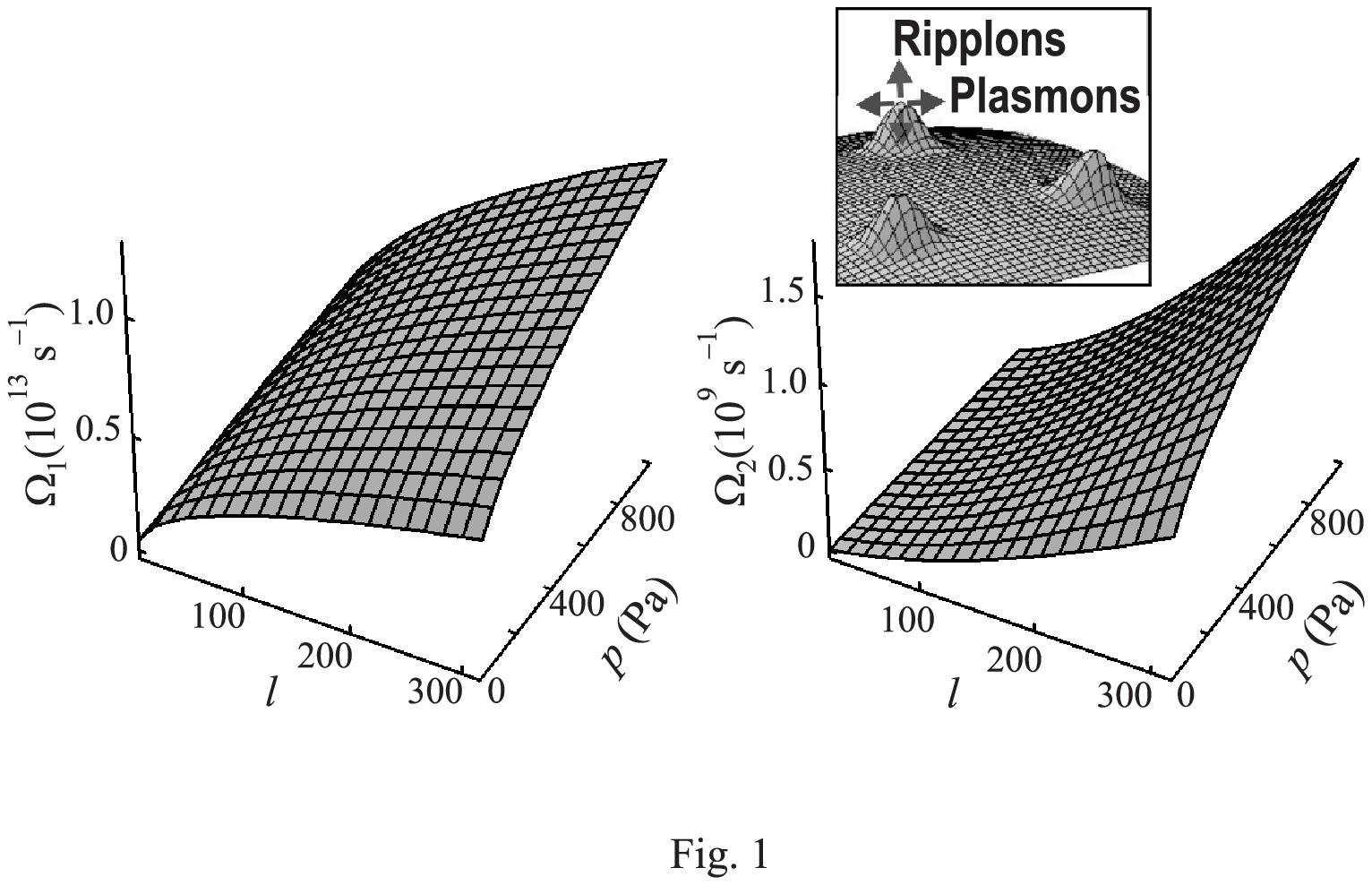}
\end{center}
\caption{Vibrational eigenfrequencies of an MEB in liquid helium,
$\Omega_{1}\left( l\right)$ (phonon modes) and $\Omega_{2}\left( l\right)$
(renormalized ripplon modes) given by Eq. (\ref{coupledmodes}) for
$N=10^{5}$ as a function of $l$ and of $p.$ Inset: a schematic picture of
the directions of motion for phonons and ripplons in an MEB. The ripplons
are excitations of the helium surface (with typical frequencies in the
MHz-GHz range), while the phonons are related to the motion of electrons
(with typical frequencies in the THz regime) tangential to the bubble
surface. Note the different scales for $\Omega_1$\ and $\Omega_2$.
}
\label{fig1}
\end{figure}

The last term in the Lagrangean (\ref{Lagrangean}) represents the
coupling between the longitudinal plasmons and the ripplons, with
coupling strength given by
\begin{equation}
\gamma _{\ell }=Ne^{2}\dfrac{(\ell +1)}{\ell +\varepsilon (\ell +1)}.
\label{gamma}
\end{equation}
For electrons on a flat helium surface, such a coupling between ripplons and
longitudinal plasmons/phonons was derived by Fisher \textit{et al.}
\cite{FisherPRL42} and detected experimentally \cite{GrimesPRL42}. For
electrons on the inner
surface of a deformed bubble, we find a similar coupling, but only ripplon
and longitudinal plasmon modes which have the same angular momentum couple
to each other.
After the diagonalisation of the ripplon-plasmon part of the Lagrangian
(\ref{Lagrangean}), we arrive at the eigenfrequencies,
\begin{equation}
\Omega _{1,2}\left( \ell \right) =\sqrt{\dfrac{1}{2}\left[ \omega
_{p}^{2}(\ell )+\omega _{r}^{2}(\ell )\right] \pm \dfrac{1}{2}\sqrt{\left[
\omega _{p}^{2}(\ell )-\omega _{r}^{2}(\ell )\right] ^{2}+4\gamma _{\ell
}^{2}}}.  \label{coupledmodes}
\end{equation}
In Fig. 1, we show the eigenfrequencies $\Omega _{1}\left( \ell \right) $
and $\Omega _{2}\left( \ell \right) $ for $N=10^{5}$ as a function of $\ell $
and $p.$ The frequency $\Omega _{2}\left( \ell \right) $ is close to the
ripplon frequency derived for an MEB within the approximation of a
conducting surface \cite{TemperePRL87}. Consequently, this frequency can be
attributed to the ripplon modes, renormalized due to the ripplon-plasmon
mixing. The admixture with the longitudinal plasmon mode is weak. The
other branch of
oscillations with the frequencies $\Omega _{2}\left( l\right) $ can be
related to the longitudinal plasmon mode admixed with a small component of
ripplon nature. In typical multielectron bubbles, the difference in
frequency between the ripplon and the plasmon modes weakens the coupling
between these modes. Unlike for electrons on a flat surface
\cite{FisherPRL42} the mixing is weak.
Finally, note that the multielectron bubble will be stable when both
$\Omega_1^2$ and $\Omega_2^2$ are positive. This condition is equivalent
to
\begin{equation}
\omega_p^2(\ell) \omega_r^2(\ell) > \gamma_{\ell}
\label{stabcond}
\end{equation}
By substituting the results (\ref{wp}), (\ref{wr}) and (\ref{gamma}) in
the inequality (\ref{stabcond}) it can easily be seen that when the
radius of the bubble is equal to the equilibrium radius, the
ripplon-plasmon mixing does not change the criterion of stability for a
MEB formulated in Ref. \cite{TemperePRL87}.

\section{Conclusions}

In this communication we studied the longitudinal plasmon-ripplon modes in
a multielectron bubble. The central result is the Lagrangean
(\ref{Lagrangean})
with the ripplon and plasmon frequencies (\ref{wr}),(\ref{wp}) and the
coupling strength (\ref{gamma}). Unlike for electrons on a flat surface, we
find that the mixing of the modes is weak for typical multielectron bubbles,
because the bare longitudinal plasmon and ripplon frequencies are
different ($\omega
_{p}(\ell )\gg \omega _{r}(\ell )$). The conditions for bubble
stability discussed in Ref. \cite{TemperePRL87} remain unaltered by the
ripplon-plasmon mixing. The analytic expressions for the longitudinal
plasmon frequencies and the coupling strength as a function of the
angular momentum allow us to derive a compact formula describing the
charge redistribution taking place on a deformed bubble. For a bubble
deformation in a particular spherical harmonic mode $\{\ell ,m\}$, the
charge will redistribute itself according to the same spherical harmonic
mode, and the amplitudes of deformation will be proportional to each
other.

\section*{Acknowledgements}
Discussions with J. Huang are gratefully
acknowledged. J. T. is supported financially by the FWO-Flanders. This
research has been supported by the Department of Energy, Grant
DE-FG02-ER45978, and by the\ GOA BOF UA 2000, IUAP, the FWO-V projects
Nos. G.0071.98, G.0306.00, G.0274.01, WOG WO.025.99N (Belgium).

\end{document}